\documentclass[11pt, a4paper]{article}

\newcommand{\ibb}[1]{\leavevmode\hbox{\kern .3em\vrule
     height 1.5ex depth .01ex width .3pt\kern-.3em\rm#1}}
\newcommand{\Ibb}[1]{ {\rm I\ifmmode\mkern
        -3.6mu\else\kern -.2em\fi#1}}

\newcommand{\C}{{\ibb C}}
\newcommand{\R}{{\Ibb R}}

\begin{document}

\centerline{\Large\bf Twisting $\kappa$-deformed phase
space\footnote{Supported by KBN grant No. 5 P03B 106
21}}\bigskip\bigskip\bigskip

\centerline{Piotr Czerhoniak} \bigskip\centerline{\it Institute of
Physics, University of Zielona G\'ora,}\centerline{\it ul.
Podg\'{o}rna 50, 65-246 Zielona G\'{o}ra, Poland}\bigskip\bigskip

\begin{abstract}
We briefly discuss the twisting procedure applied to the
$\kappa$-defor\-med space-time. It appears that one can consider
only two kinds of such twistings: in space and time directions.
For both types of twisitngs we introduce related phase spaces and
consider briefly their properties. We discuss in detail the
changes of duality relations under the action of twist. The
Jordanian twisted space-time and phase space in D=2 are also
commented.
\end{abstract}




\section{Introduction}
Recently, some suggestions appeared that the classical Lorentz
invariance should be treated as an approximate symmetry in
ultra-high energy processes (shift of GZK kinematic threshold) and
the relativistic space-time symmetries on Planck scale should be
modified \cite{COM1}. There are also some theoretical predictions
coming from the string theory and quantum gravity models that
space-time at very short distances of order Planck length should
be quantum i.e. noncommutative \cite{DFR}. One can modify the
standard Lorentz or Poincar\'{e} relativistic symmetry in two
different ways: obtaining a commutative space-time (as in the
standard relativistic theory) or noncommutative space-time with
space and time non-commuting variables.\medskip\\The first type of
modified relativistic symmetries follows from the concept of the
double special relativistic (DSR) theory introduced by
Amelino-Camelia \cite{GAC1}, where two observer-independent
parameters (scales) -- $c , \lambda_p$ -- velocity of light and
Planck length play the fundamental role. In this framework, two
basic models of such DSR theory are considered: DSR1 theory
proposed by Amelino-Camelia and DSR2 theory considered by Magueijo
and Smolin \cite{MS}. In both proposals the energy-momentum vector
space is extended to the four-momentum algebra considered as
enveloping algebra of this linear energy-momentum space. In this
algebra the nonlinear transformation of the basis vectors (of
energy-momentum space) is realized according to DSR1 or DSR2
theory models. In this way we get four non-linearly transformed
generators belonging to the energy-momentum algebra which are
assumed to represent physical energy and momentum. Using this
transformed four-momentum basis we obtain for instance the
standard dispersion relation in a modified form which is, however,
quite equivalent (under an inverse nonlinear transformation) to
the standard one. One can also consider the action of the
relativistic symmetry on physical energy and momentum as a
nonlinear realization of the Lorentz group and express the
addition laws of energy and momentum by the coproduct
(non-linearly transformed trivial coproduct) \cite{LN} and
extending in this way the energy-momentum algebra to a bialgebra
structure. Space-time related to this bialgebra structure is
defined as a dual bialgebra where  multiplication of the
space-time variables follows from the coproduct. The momentum
coproduct is symmetric, therefore space-time is commuting. In
consistency with conclusions in \cite{LN}, recently it has been
observed \cite{AT} that DSR theories with symmetric coproduct law
describe just the standard special relativity framework in
nonlinear disguise.\medskip\\The second type of modified
relativistic symmetries bases on the concept of the Hopf algebra
(quantum group) \cite{ABE}, \cite{DR1} and quantum deformations of
the classical Lorentz or Poincar\'{e} symmetry \cite{M}. In this
approach there is a distinguished deformation, the so called
$\kappa$-deformation of the Poincar\'{e} symmetry \cite{LNR},
where $O(3)$-rotational symmetry is not deformed. The Hopf algebra
structure of this deformation will be discussed in detail in
Section 2. However, we would like to stress that the four-momentum
algebra for DSR1 theory and the $\kappa$-deformed theory (in
bicrossproduct basis) are the same, and give us the same physical
predictions. The main difference between both theories lies in
their coalgebra sectors, because in the $\kappa$-deformed
Poincar\'{e} algebra the coproduct of the four-momentum is
non-symmetric. This fact has important consequences for a
possibility of physical interpretation of the $\kappa$-deformed
addition law of energy-momentum. On the other hand, this
non-symmetricity of coproduct allows us to obtain
non-commutativity of space-time in the Hopf algebra framework. The
idea of a non-commutativity of space-time appeared in physics
since Snyder's articles \cite{Snyd}, where the space-time is
isomorphic to a quotient of de Sitter $SO(4,1)$ and Lorentz
$SO(3,1)$ groups.\medskip\\Further, we shall consider
$\kappa$-deformed space-time defined as a dual Hopf algebra to the
$\kappa$-deformed four-momentum algebra considered as a subalgebra
of $\kappa$-Poincar\'{e} Hopf algebra i.e. the space-time is
understood in Majid and Ruegg sense \cite{MR}. Such a space-time
has a Hopf algebra structure given by coproduct of trivial form.
In this paper we shall consider a twisting operation acting on the
$\kappa$-deformed space-time algebra. The notion of twisting of
Hopf algebra was introduced by Drinfeld \cite{DR} and then applied
to enveloping algebras of simple Lie algebras \cite{RE} and the
applications to physical symmetry algebras one can find in
\cite{APL} and recently in $\kappa$-deformations of Weyl and
conformal symmetries \cite{LLM}.\\For a given pair of space-time
and four-momentum algebras one can define a phase-space algebra
using the notion of cross product algebra \cite{M}. It is well
known that although both paired algebras have a Hopf algebra
structure, the phase space algebra is no longer a Hopf algebra.
The case of $\kappa$-deformed phase space is considered in
\cite{LN1}, \cite{ALN}.\medskip\\In Section 2 we recall the Hopf
algebra structure of $\kappa$-Poincar\'{e} algebra in a very
useful bicrossproduct basis \cite{MR} in which the Lorentz algebra
sector has a classical form. We also discuss the $\kappa$-deformed
space-time underlying the role of duality relations.\medskip\\In
Section 3.1 we show that one can define only two kinds of twisting
operations on $\kappa$-deformed space-time: twisting in space
directions (SD) and in time direction (TD). For both cases we
derive their Hopf algebra structure. In Section 3.2 we extend our
dual pair to a phase space algebra and we find phase space
commutation relations. It appears that the SD-twisted phase space
can not be considered as a linear vector space because of
nonlinear phase space commutation relations.\medskip\\ Section 4
is devoted to the derivation of duality relations between the
twisted space-time algebra and the momentum algebra for simplicity
in two dimensional space-time. It is more a computational part of
the paper where we find that the space-time twisting effectively
changes only duality relations. In this way the basis of twisted
space-time algebra becomes non-orthogonal (one can say "twisted")
with respect to the momentum basis. We also consider D=2 twisted
$\kappa$-deformed space-time by nonsymmetric twisting function
(Jordanian twist) \cite{Ogiev}, \cite{Jord} and dual momentum
algebra. The phase space for this Jordanian deformation is
obtained and considered.

\section{$\kappa$-Poincar\'{e} algebra in the bicrossproduct
basis \\and $\kappa$-deformed space-time}

Let us recall the structure of $\kappa$-Poincar\'{e} algebra given
in the bicrossproduct basis \cite{MR}\medskip\\- non-deformed
({\it classical}) Lorentz algebra
($g_{\mu\nu}=(1,-1,-1,-1)$)\begin{equation} [M_{\mu\nu},
M_{\rho\tau}] = - i(g_{\mu\rho}M_{\nu\tau}+
g_{\nu\tau}M_{\mu\rho}-g_{\mu\tau}M_{\nu\rho}
-g_{\nu\rho}M_{\mu\tau})\, ,\label{r21}\end{equation}-  deformed
covariance relations ($M_i= \frac12\epsilon_{ijk} M_{jk}, \
N_i=M_{i0}$)
\begin{equation}
\begin{array}{lll}
[ M_i , p_j ] & = i\epsilon_{ijk}p_k\, ,\qquad & [ M_i , p_0 ]  =
0\, ,\medskip\cr [ N_i , p_0 ] & = i p_i\, , \qquad & [ p_\mu ,
p_\nu ] = 0\, ,\medskip\cr[ N_i , p_j ]& =
i\delta_{ij}\left[\kappa c\sinh(\frac {p_0}{\kappa c})
e^{-\frac{p_0}{\kappa c}}  + \frac1{2\kappa c} (\vec p)^2\right] -
\frac {i}{\kappa c} p_i p_j\,
,&\end{array}\label{r22}\end{equation}where \quad $\kappa$ -
massive deformation parameter and\quad c - light velocity.\\One
can extend this algebra to a Hopf algebra defining the coalgebra
sector by {\it coproduct} $\Delta(X)$
\begin{equation}
\begin{array}{ll}
\Delta(M_i) = & M_i\otimes 1 + 1\otimes M_i\medskip\, ,\cr
\Delta(N_i) = & N_i\otimes 1 + e^{-\frac{p_0} {\kappa c}}\otimes
N_i + {\frac{1}{\kappa c}}\epsilon_{ijk}p_j\otimes M_k\, ,
\medskip\cr \Delta(p_0) = & p_0\otimes 1 + 1\otimes p_0\, ,\medskip\cr
\Delta(p_i) = & p_i\otimes 1 + e^{-{\frac{p_0}{\kappa c}}}\otimes
p_i\, ,\end{array}\label{r23}\end{equation}with {\it antipode}
$S(X)$ and {\it counit} $\epsilon(X)$
\begin{equation}\begin{array}{lll}S(M_i) & = -M_i\, ,\qquad & S(N_i)
 = -e^{\frac{p_0}{\kappa c}}N_i+{\frac{1}{\kappa
c}}\epsilon_{ijk}e^{\frac{p_0}{\kappa c}}p_j M_k\, ,\medskip\cr
S(p_0)& = -p_0\, , & S(p_i) = -p_i e^{\frac{p_0}{\kappa c}}\, ,
\end{array}\label{r24}\end{equation} \begin{equation} \epsilon(X) \ = \
0\, ,\quad \epsilon(1) \ = \ 1\, ,\qquad for \qquad X= M_i, N_i,
p_\mu\, .\label{r25}\end{equation}Therefore, in the
$\kappa$-Poincar\'{e} Hopf algebra one can distinguish the
following three subalgebras\\- classical Lorentz algebra $\cal{L}
= \{M_{\mu \nu}\}$ given by (\ref{r21}),\\- non-deformed
$O(3)$-rotation algebra ${\cal{M}} = \{M_i , \Delta , S ,
\epsilon\}$ as a Hopf subalgebra with trivial coproduct
$\Delta(M_i)$,\\- abelian four-momentum algebra ${\cal{P}}_\kappa
= \{p_\mu , \Delta , S , \epsilon \}$ as a Hopf subalgebra
with\newline non-symmetric coproduct $\Delta(p_\mu)$ given by the
relations\begin{equation}\begin{array}{lll}[ p_\mu , p_\nu ] & = \
0\, ,\qquad & \epsilon( p_\mu ) \ = \ 0\, ,\medskip\cr \Delta(p_0)
& = \ p_0\otimes 1 + 1\otimes p_0\, , & \Delta( p_i ) \ = \
p_i\otimes 1 + e^{-\frac{p_0}{\kappa c}}\otimes p_i\, ,\medskip\cr
S( p_0 ) & = \ -p_0\, , & S(p_i) \ = \ -p_i e^{\frac{p_0}{\kappa
c}}\, .\end{array}\label{r26}\end{equation}The Hopf four-momentum
algebra ${\cal{P}}_\kappa$ can be considered as a deformation of
universal enveloping algebra of classical translation algebra,
generated (in Drinfel'd sense \cite{DR1}) by polynomial functions
of linear momentum generators $p_\mu$ (translations).\\To define
the $\kappa$-deformed space-time we adopt Majid and Ruegg point of
view \cite{MR}, namely, using this algebra ${\cal{P}}_\kappa$ we
define $\kappa$-deformed space-time algebra ${\cal{X}}_\kappa$ in
a natural way, as a dual algebra to four-momentum translation
algebra ${\cal{P}}_\kappa$. One can assume the standard duality
relations between linear bases of both algebras, respecting Hopf
algebra structure.\\Let ${\cal{X}}_\kappa$ be generated by
space-time variables $ \{x_0 , \vec{x} \}$ dual to four-momentum
$p_\mu$ satisfying the following duality relations
\begin{eqnarray}&< p_\mu , x_\nu > = i\,g_{\mu\nu}\,
,\qquad\qquad & g_{\mu \nu} = (1 , -1 , -1 , -1 )\,
,\label{r27a}\medskip\\&< p_\mu , 1 > = < 1 , x_\mu > = 0\,
,\quad& < 1, 1 > = 1\, ,\label{r27b}\medskip\\&< p\, q , x > = < p
\otimes q , \Delta(x) >\, , & < p , x\, y > = < \Delta(p) ,
x\otimes y >\, ,\label{r27c}\end{eqnarray}where $p_\mu, p , q\in
{\cal{P}}_\kappa\, ,\quad x_\nu , x , y\in {\cal{X}}_\kappa$,\\or
using Sweedler coproduct notation (see for instance \cite{M})
$\Delta(a) = a_{(1)}\otimes a_{(2)}$ relations (\ref{r27c}) can be
rewritten as\begin{eqnarray}< p\, q , x > &=& < p \otimes q ,
\Delta(x) > \ = \ < p , x_{(1)} > < q , x_{(2)}
>\, ,\label{r28a}\medskip\\< p , x\, y > &=& < \Delta(p) ,
 x\otimes y > \ = \ < p_{(1)} , x > < p_{(2)} , y >\, .
\label{r28b}\end{eqnarray}Relations (6-9) do not describe the Hopf
algebra structure of space-time algebra uniquely, in particular
they describe a wide class of coproducts $\Delta(x_\nu)$. The form
of space-time coproduct depends on the choice of higher order
duality relations and vice versa. If we assume orthogonality of
space-time and four-momentum monomial basis in the form ($i , j =
1, 2, 3,\quad k , l , m , n = 0, 1, 2, \dots$)\begin{equation} <
{p_i}^m\,{p_0}^n , {x_j}^k\,{x_0}^l
> \ = \ m!\,n!\,\delta_{m k} \delta_{n l} \delta_{i j} < p_0 ,x_0
>^n < p_i , x_i >^m\, ,\label{r29}\end{equation}then we get the
following form of non-commuting space-time Hopf algebra \cite{MR}
\begin{eqnarray} [ x_0 , x_i ]& = \ < e^{-\frac{p_0}{\kappa c}} ,
x_0 > x_i \ = \ -\frac{i}{\kappa c} x_i\, ,\qquad\qquad\quad & [
x_i , x_j ] \ = \ 0\, ,\label{r210a}\bigskip\\ \Delta_0(x_\mu)& =
\ x_\mu\otimes 1 + 1\otimes x_\mu\, ,\quad S(x_\mu) \ = \ -
x_\mu\, ,\quad & \epsilon(x_\mu) \ = \ 0\,
,\label{r210b}\end{eqnarray} with trivial, symmetric coproduct
$\Delta_0(x_\mu)$.\\In this place we would like to stress the
difference between the notion of space-time algebra and the
concept of deformed (quantum or $\kappa$-deformed) space-time --
the linear span of space-time variables $x_\mu$. For the commuting
(classical) space-time both notions are equivalent -- commuting
space-time algebra is simply the algebra of commuting (polynomial)
functions on space-time. Classical space-time can be regarded as a
dual to the four-momentum (translation) space because of trivial
momentum coproduct. However, in the deformed (quantum) case the
momentum coproduct (\ref{r26}) contains the exponential factor
which belongs to the momentum algebra. It is the reason why we use
the notion of space-time algebra for non-commuting
space-time.\medskip\\Because the space-time variables $x_\mu$ do
not commute among themselves, one can also choose a space-time
monomial basis with opposite ordering and satisfying the duality
relations \cite{LN1}\begin{eqnarray}< {p_i}^m\,{p_0}^n ,
{x_0}^l\,{x_j}^k
> &=& \frac{l!\,\delta_{k m} \delta_{i j}}{(l-n)!}< p_i , x_i >^m
< p_0 , x_0 >^n < e^{-\frac{m p_0}{\kappa c}} , x_0^{l-n}
>\, ,\bigskip\cr < {p_i}^m\,{p_0}^n , {x_0}^l\,{x_j}^k > &=&
0\qquad\qquad for\qquad n > l\,
,\label{r211}\end{eqnarray}therefore, for this ordering, we obtain
non-orthogonal duality relations for $0\leq n \leq l$. Both
relations (\ref{r29}) and (\ref{r211}) one can rewrite in a more
convenient form using the exponential generating function
\begin{eqnarray}< {p_i}^m\,e^{\xi p_0} ,\, {x_j}^k\, x_0^l
> &=& < e^{\xi p_0} ,\, x_0^l > < {p_i}^m , {x_j}^k >\, ,\label{r212a}
\medskip\\ < {p_i}^m\,e^{\xi p_0} , \,x_0^l\, {x_j}^k > &=&
 < e^{\xi p_0 - \frac{m p_0}{\kappa}} ,\, x_0^l > < {p_i}^m ,\,
{x_j}^k > =\nonumber\medskip\\&=& < e^{\xi p_0} , (x_0 -
i\frac{k}{\kappa c} )^l > < {p_i}^m ,\, {x_j}^k >\,
.\label{r212b}\end{eqnarray}Comparing these formulae one can easy
obtain the general form of the space-time commutation
relations\begin{equation}[ {x_i}^k , {x_0}^l ] \ = \ {x_i}^k
\left\{ {x_0}^l - \left( x_0 - i\frac{k}{\kappa
c}\right)^l\right\}\, ,\label{r213}\end{equation}in the
right-time-ordered basis (\ref{r29}) (with all powers of the time
variable $x_0$ on the right).\\We would like to notice, that the
duality relations (\ref{r29}) have the same form as in the case of
the classical Poincar\'{e} algebra with trivial Hopf structure and
its commuting dual space-time algebra. This duality relation can
be rewritten in an equivalent form (see
\cite{MR})\begin{equation}< f(p_i , p_0) , :\phi( x_j , x_0 ): > \
= \ \left( f\left(-i\frac{\partial}{\partial x_i} , i
\frac{\partial}{\partial x_0}\right)\, \phi\right) ( 0 , 0 )\,
,\label{r214}\end{equation}for polynomial functions $ f , \phi $
\, and $:\phi:$ denotes right-time-ordered polynomial.\\Further,
we shall consider the coproducts of space-time variables related
to $\Delta_0(x_\mu)$ by twisting procedure, and we shall discuss
twisted duality relations.

\section{Twisted $\kappa$-deformed space-time and phase
space}
\subsection{Twisting procedure}It is well known, that
$\kappa$-deformed space-time (\ref{r210a}) can be extended to a
Hopf algebra up to similarity transformation (twisting) in the
coalgebra sector. The choice of coproduct in the form
(\ref{r210b}) is the simplest one, however one can also consider a
more general class of twisted coproducts \cite{RE}, \cite{APL}
given by the following similarity transformation
\begin{equation}\Delta^F( x_\mu ) \ = \ F \Delta_0( x_\mu )
F^{-1}\, ,\label{r31}\end{equation}where $(a\otimes b)(c\otimes d)
= ac\otimes b d $ and an invertible twisting function $F \in
{\cal{X}}_\kappa\otimes {\cal{X}}_\kappa$ satisfies the additional
Hopf structure requirements which follow from the properties of
twisted coproduct $\Delta^F( x_\mu)$\medskip\\{\it
coassociativity}\begin{equation}(\Delta^F\otimes 1) \Delta^F
(x_\mu) \ = \ (1\otimes \Delta^F ) \Delta^F (x_\mu)\,
,\label{r33}\end{equation}{\it consistency
relations}\begin{eqnarray}(\epsilon\otimes 1)\circ \Delta^F( x_\mu
) &=& (1\otimes \epsilon)\circ \Delta^F( x_\mu ) \ = \ x_\mu\,
,\label{r34}\medskip\\( S^F\otimes 1 )\circ \Delta^F( x_\mu ) &=&
(1\otimes S^F)\circ \Delta^F( x_\mu ) \ = \ 0\,
,\label{r35}\end{eqnarray}where we denote $(a\otimes b)\circ
(c\otimes d) = a c b d $. We would like to notice, that twisted
space-time is defined by twisted coproduct $\Delta^F$  and
antipode $S^F$.\bigskip\\The coassociativity condition (\ref{r33})
can be rewritten in a more familiar form as a 2-cocycle condition
imposed on the twisting function $F$ \cite{M}
\begin{equation}(1\otimes F) ( 1\otimes \Delta_0 ) F \ = \ (
F\otimes 1) ( \Delta_0\otimes 1 ) F\,
.\label{r352}\end{equation}Without loss of generality we assume
the following exponential form \cite{RE} of the twisting
function\begin{equation} F \ = \ \exp\left(\sum\,\phi_n\otimes
\phi^n\right)\, ,\label{r353}\end{equation}where $\phi_n , \phi^n
\in {\cal{X}}_\kappa$. Then the coassociativity condition
(\ref{r352}) one can express by the formula\begin{equation}
e^{1\otimes \phi_n\otimes \phi^n} e^{\phi_n\otimes
\Delta_0(\phi^n)} \ = \ e^{\phi_n\otimes \phi^n\otimes 1}
e^{\Delta_0(\phi_n)\otimes \phi^n}\, .\label{r354}\end{equation}
Let us notice that the relations (\ref{r210a}) - (\ref{r210b})
define noncommuting space-time as a four dimensional Lie algebra
and  $\kappa$-deformed space-time algebra as an enveloping Lie
algebra  ${\cal{X}}_\kappa$. Therefore, one can use the results of
ref. \cite{RE} to find twisted space-time as a twisted Lie algebra
with the Hopf structure. It is equivalent to put the additional
requirements on the twisting function $F$
\begin{equation}(\Delta_0\otimes 1) F \ = \ F_{1 3} F_{2 3} \ = \
F_{2 3} F_{1 3}\, ,\quad [ F_{1 2}\,, F_{2 3} ] \ = \ 0\,
,\label{r3d1}\end{equation}where we use the standard notation
\begin{equation}F_{1 2} \ = \ F\otimes 1\, ,\quad F_{2 3} \ = \
1\otimes F\, ,\quad F_{1 3} \ = \ e^{\phi_n\otimes 1\otimes
\phi^n}\, .\label{r3d2}\end{equation} In this framework, the
elements $\phi_n\,,\phi^n$ belong to commutative subalgebra of
${\cal{X}}_\kappa$, with trivial coproduct
\begin{eqnarray}\Delta_0 ( \phi^n ) & =& \phi^n \otimes 1 +
1\otimes \phi^n\, ,\qquad [ \phi_n\otimes \phi^n\, , \Delta_0(
\phi^m ) ] \ = \ 0\, ,\nonumber\\\Delta_0 ( \phi_n ) & =& \phi_n
\otimes 1 + 1\otimes \phi_n\, ,\qquad [ \phi_n\otimes \phi^n\, ,
\Delta_0( \phi_m ) ] \ = \ 0\, .\label{r355}\end{eqnarray} Of
course, it is not the most general case of a twisting function
$F$. For instance, if we do not assume commutativity of $F_{12},
F_{23}$ in (\ref{r3d1}) (so, we give up a Lie algebra twisting
framework) we get
\begin{eqnarray}\Delta_0 ( \phi_n ) & =& \phi_n \otimes 1 +
1\otimes \phi_n\, ,\qquad [ \phi_m\, , \phi_n ] \ = \ 0\, , \\
\Delta^F ( \phi^n ) & =& \phi^n \otimes 1 + 1\otimes \phi^n\,
,\qquad [ \phi^m\, , \phi^n ] \ = \ 0\,
.\label{r3d3}\end{eqnarray} In this case we obtain more general
twisting function $F$ of noncommuting space-time (so called
Jordanian twist, discussed in Section 4).
\bigskip\\If we consider the twisting of $\kappa$-deformed space
time algebra, we deal with symmetric function $F$ because the dual
fourmomentum algebra is commuting one. Therefore, a symmetric
twisting function $F$ satisfying the relations (\ref{r352}) and
(\ref{r3d1}) is given by
\begin{equation}F \ = \ e^{\phi\otimes \phi} \ = \ \exp[(a
x_0 + b_i x_i)\otimes (a x_0 + b_j x_j)]\,
,\label{r3d4}\end{equation}where the four twisting parameters $a ,
b_i$ are in general the complex numbers. For this twisting
function $F$ we immediately obtain the following formulae for
twisted coproduct of space-time variables
\begin{eqnarray}\Delta^F(x_0) &=& \Delta_0(x_0) + \frac{b_i}{a}
\left[x_i\otimes \left(1 - e^{a \lambda \phi}\right) + \left(1 -
e^{a \lambda \phi}\right)\otimes x_i\right]\, ,\label{r3d5}\\
\Delta^F(x_i) &=& \Delta_0(x_i) + x_i\otimes \left(e^{a \lambda
\phi} - 1\right) + \left(e^{a \lambda \phi} - 1\right)\otimes
x_i\, ,\label{r3d6}\end{eqnarray}where $\lambda = -i/\kappa c$
(\ref{r210a}).\bigskip\\Because a time variable should not depend
on the space inversion, therefore it is physically reasonable to
assume that the twisting function $F$ is a space-inversion
invariant. Using this fact one can construct two twisting
functions related with two abelian subalgebras of
$\kappa$-deformed space-time. The first one generated by the time
variable $x_0$ and the other one by the space variables $x_i$
(\ref{r210a}), both with trivial coproduct (\ref{r210b})\\{\it
twisting of space directions} (SD)\begin{equation}F_0( a ) \ = \
e^{a\,x_0\otimes x_0}\, ,\label{r36}\end{equation}{\it twisting of
time direction} (TD)\begin{equation}F( b ) \ = \ e^{b_{i
j}\,x_i\otimes x_j}\, ,\label{r37}\end{equation}where $a,\,
b_{ij}=b_i b_j\in \C$ in general case. If we consider the
space-time Hopf algebra with involution $*$ satisfying $(a\otimes
b)^*=a^*\otimes b^*$ then, the assumption of hermiticity of
space-time generators $x^*_\mu = x_\mu$ implies the unitarity of
twisting functions i.e. $a,\, b_{ij}\in i\,\R$ are pure imaginary
complex numbers.

\noindent For the twisting function $F_0(a)$, using formulae
(\ref{r31}) and (\ref{r35}) we obtain SD-twisted space-time Hopf
algebra ${\cal{X}}_\kappa(\alpha)$ hermitian basis)
\begin{eqnarray} [ x_0 , x_i ] &=& -\frac{i}{\kappa c} x_i\, ,
\qquad\qquad  [ x_i , x_j ] \ = \ 0\,
,\nonumber\bigskip\\\Delta^{F_0(a)}(x_0) &=& \Delta_{\alpha}( x_0
) \ = \ x_0\otimes 1 + 1\otimes x_0\, ,\nonumber\bigskip\\
\Delta^{F_0(a)}(x_i) &=& \Delta_{\alpha}( x_i ) \ = \ x_i\otimes
e^{\alpha x_0} + e^{\alpha x_0}\otimes x_i\,
,\label{r38}\bigskip\\ S^{F_0(a)}( x_i) &=& S_\alpha( x_i) \ = \ -
x_i e^{\alpha(\frac{i}{\kappa c} - 2 x_0)}\, ,\nonumber\bigskip\\
S^{F_0(a)}( x_0) &=& S_\alpha( x_0 ) \ = \ - x_0\,
,\qquad\qquad\quad \epsilon( x_\mu ) \ = \ 0\,
,\nonumber\end{eqnarray}where\begin{equation}\alpha\equiv a
<e^{-\frac{p_0}{\kappa c}} , x_0 > \ = \ -\frac{i\,a}{\kappa c}\in
\R\, ,\label{r39}\end{equation}and similarly, choosing the
twisting function as $F(b)$ we get the TD-twisted space-time
Hopf-algebra ${\cal{X}}_\kappa(\beta)$ hermitian basis)
\begin{eqnarray} [ x_0 , x_i ] &=& -\frac{i}{\kappa c} x_i\, ,
\qquad\qquad  [ x_i , x_j ] \ = \ 0\, ,\nonumber\bigskip\\
\Delta^{F(b)}(x_0) &=& \Delta_{\beta}( x_0 ) \ = \ x_0\otimes 1 +
1\otimes x_0 + \beta_{i j}\,x_i\otimes x_j\, ,\nonumber\bigskip\\
\Delta^{F(b)}(x_i) &=& \Delta_{\beta}( x_i ) \ = \ x_i\otimes 1 +
1\otimes x_i\, ,\label{r310}\bigskip\\ S^{F(b)}( x_i) &=& S_\beta(
x_i) \ = \ - x_i \, ,\nonumber\bigskip\\ S^{F(b)}( x_0) &=&
S_\beta( x_0) \ = \ - x_0 + \beta_{i j}\,x_i x_j\,
,\qquad\qquad\quad \epsilon( x_\mu ) \ = \ 0\, ,\nonumber
\end{eqnarray} where\begin{equation}\beta_{i j} \ = \ \frac{2
i}{\kappa c}\,b_{i j}\in \R\, .\label{r311}\end{equation} It is
obvious, that in the limit $\alpha,\,\beta\rightarrow 0$ we obtain
a $\kappa$-deformed space-time algebra given by
(\ref{r210a}-\ref{r210b}).\bigskip\\It is well known that a
twisting by a symmetric function $F$ is equivalent to nonlinear
transformation of space-time, therefore SD-twisted coproduct
(\ref{r38}) is given by the function $s(x_\mu)$
\begin{eqnarray}\Delta_\alpha(x_0) &=& \Delta_0(s(x_0))\, ,\qquad
s(x_0) \ = \ x_0\, ,\nonumber\\ \Delta_\alpha(x_i) &=&
\Delta_0(s(x_i))\, ,\qquad s(x_i) \ = \ x_i e^{\alpha x_0}\,
,\nonumber\\\Delta_0(s(x_i)) &=& s(x_i)\otimes e^{\alpha s(x_0)} +
e^{\alpha s(x_0)}\otimes s(x_i)\, ,\label{r3d7}\end{eqnarray}and
analogously, TD-twisted coproduct (\ref{r310}) can be rewritten
using function $t(x_\mu)$
\begin{eqnarray}\Delta_\beta(x_0) &=& \Delta_0(t(x_0))\, ,\qquad
t(x_0) \ = \ x_0 + \frac{1}{2} \beta_{i j} x_i x_j\, ,\nonumber\\
\Delta_\beta(x_i) &=& \Delta_0(t(x_i))\, ,\qquad t(x_i) \ = \
x_i\, ,\nonumber\\\Delta_0(t(x_0)) &=& t(x_0)\otimes 1 + 1\otimes
t(x_0) + \beta_{i j} t(x_i)\otimes t(x_j)\,
.\label{r3d8}\end{eqnarray} We see that the twisting of $\kappa$-
deformed space-time is equivalent to nonlinear transformation of
the space-time variables.

\subsection{Phase space as cross product algebra}Let us notice,
that we have two pairs of dual Hopf algebras
${\cal{X}}_\kappa(\alpha)\otimes {\cal{P}}_\kappa$ and
${\cal{X}}_\kappa(\beta)\otimes {\cal{P}}_\kappa$ which in the
non-deformed limit $\kappa\to\infty$ turn out to be classical
space-time and momentum algebras with multiplication defined by
the commutator, therefore we should get the quantum-mechanical
phase space with standard Heisenberg commutation relations.\\In
order to construct such a deformed phase space algebra
$\Pi_\kappa$ isomorphic as a vector space to $\Pi_\kappa \sim
{\cal{X}}_\kappa\otimes {\cal{P}}_\kappa$ one has to extend the
commutation relations (\ref{r26}) and (\ref{r38}) or (\ref{r310})
by adding a cross commutators between ${\cal{X}}_\kappa$ and
${\cal{P}}_\kappa$.\\It appears that a consistent construction of
phase space $\Pi_\kappa$ can be done using the notion of a left
(right) cross product (smash product) algebra \cite{M}. For
simplicity, we shall consider only left cross product
algebra.\medskip\\One can define a {\it left action}
(representation) of the momentum algebra ${\cal{P}}_\kappa$ on the
space-time algebra ${\cal{X}}_\kappa$ as a linear map
\begin{equation}\triangleright :\, {\cal{P}}_\kappa\otimes
{\cal{X}}_\kappa\rightarrow {\cal{X}}_\kappa :\, p\otimes
x\rightarrow p\triangleright x\, ,\label{r313}\end{equation}such
that\begin{equation} (p\tilde{p})\triangleright x \ = \
p\triangleright (\tilde{p} \triangleright x)\, ,\quad
1\triangleright x \ = \ x\, .\label{r314}\end{equation}We choose
the following left action\begin{equation} p\triangleright x \ = \
x_{(1)} < p\,, x_{(2)} >\, ,\label{r315}\end{equation} therefore
${\cal{X}}_\kappa$ is a left ${\cal{P}}_\kappa$-module or even a
left ${\cal{P}}_\kappa$-module algebra, because ${\cal{X}}_\kappa$
and ${\cal{P}}_\kappa$ are also Hopf algebras and the left action
(\ref{r315}) satisfies\begin{equation} p\triangleright
(x\tilde{x}) \ = \ (p_{(1)}\triangleright x) (p_{(2)}
\triangleright \tilde{x})\, ,\quad p\triangleright 1=\epsilon(p)
1\, .\label{r316}\end{equation}This implies that we can regard the
twisted space-time as the left $\kappa$-deformed momentum
${\cal{P}}_\kappa$-module algebra for both choices of twisting
functions (\ref{r36}) and (\ref{r37}) with the following left
action (\ref{r315}) in the case of SD-twisted space-time
(\ref{r38})\begin{eqnarray} p_0 \triangleright x_0 & = \ i\,
,\qquad\qquad & p_i \triangleright x_0 \ = \ 0\, ,\cr p_0
\triangleright x_i & = \ i\,\alpha x_i\, ,\qquad\,\, & p_i
\triangleright x_j \ = \ -i\,\delta_{i j} e^{\alpha x_0}\, ,
\label{r317}\end{eqnarray}and in the case of TD-twisted space-time
(\ref{r310}) we get\begin{eqnarray} p_0 \triangleright x_0 & = \
i\, ,\qquad & p_k \triangleright x_0 \ = \ -\,i \beta_{i k} x_i\,
,\cr p_0 \triangleright x_i & = \ 0\, ,\qquad\,\, & p_k
\triangleright x_i \ = \ -i\,\delta_{k i} \,
.\label{r318}\end{eqnarray}We recall the definition of a {\it left
cross product algebra} \cite{M}.\medskip\\Let ${\cal{P}}_\kappa$
be a Hopf algebra and ${\cal{X}}_\kappa$ a left
${\cal{P}}_\kappa$-module algebra. A left {\it cross product
algebra} $\Pi_\kappa = {\cal{X}}_\kappa
>\!\!\!\triangleleft\,{\cal{P}}_\kappa$ is a vector space ${\cal{X}}_\kappa\otimes
{\cal{P}}_\kappa$ with the product ({\it left cross
product})\begin{equation}(x\otimes p)(\tilde{x}\otimes \tilde{p})
\ = \ x(p_{(1)}\triangleright \tilde{x})\otimes p_{(2)}\tilde{p}\,
,\label{r319}\end{equation}and the unit element $1\otimes 1$,
where $x , \tilde{x} \in {\cal{X}}_\kappa$ and $p , \tilde{p} \in
{\cal{P}}_\kappa$. It appears that $\Pi_\kappa = {\cal{X}}_\kappa
>\!\!\!\triangleleft\,{\cal{P}}_\kappa$ is an associative algebra,
however it can not be extended to a Hopf algebra (\cite{M}). This
fact suggests that one can construct $\kappa$-deformed phase-space
$\Pi$ for many particle system in a usual way as a tensor product
algebra $\Pi = \Pi_\kappa\otimes \cdots \otimes \Pi_\kappa$.
\medskip\\The obvious isomorphism ${\cal{X}}_\kappa\sim
{\cal{X}}_\kappa\otimes 1$, ${\cal{P}}_\kappa\sim 1\otimes
{\cal{P}}_\kappa$ allows us to define the commutator for the whole
phase space $\Pi_\kappa$
\begin{equation}[x,p] \ = \ x\circ p-p\circ x\, ,
\qquad  x\circ p \ = \ x\otimes p\, , \qquad p\circ
x=(p_{(1)}\triangleright x)\otimes p_{(2)}\,
.\label{r320}\end{equation}Using this definition and formulae
(\ref{r38}) and (\ref{r317}) we obtain the commutation relations
for the linear basis of SD-{\it twisted phase space}
$\Pi_\kappa(\alpha) = {\cal{X}}_\kappa(\alpha)
>\!\!\!\triangleleft\,{\cal{P}}_\kappa$\begin{equation}
\begin{array}{lll} [ x_0 , x_i ] \ = & -\frac{i}{\kappa c} x_i\, ,
\qquad\qquad & [ x_i , x_j ] \ = \ 0\, ,\medskip\cr [ x_i , p_0 ]
\ = & -i\,\alpha x_i\, , & [ x_0 , p_i ] \ = \ \frac{i}{\kappa c}
p_i\, ,\medskip\cr [ x_0 , p_0 ] \ = & - i\, , & [ p_\mu , p_\nu ]
\ = \ 0\, ,\bigskip\cr [ x_i , p_j ] \ = & i\,\delta_{i j}
e^{\alpha x_0} + \left(1 - e^{-\frac{i \alpha}{\kappa c}}\right)
x_i\,p_j& \, .\end{array}\label{r321}\end{equation}Let us notice
that because of the exponential term in the last relation, the
SD-twisted phase space can be considered only as an algebra.\\In
the limit $\alpha \to 0$ we obtain the standard $\kappa$-deformed
phase space considered in \cite{LN1}, a deformed generalization of
Heisenberg algebra. It is interesting to notice that one can also
consider the limit $\kappa\to\infty , \alpha=const.$ (i.e. one can
assume the linear dependence of the twisting parameter $a$ on the
deformation parameter $\kappa$, see (\ref{r39})) of phase space
$\Pi_\kappa(\alpha) \to \Pi_\infty(\alpha)$ given by the
non-vanishing commutators\begin{equation}[ x_0 , p_0 ] \ = \ -i\,
,\quad [ x_i , p_j ] \ = \ i\,\delta_{i j} e^{\alpha x_0}\, ,\quad
[ x_i , p_0 ] \ = \ -i\,\alpha x_i\,
,\label{r322}\end{equation}with commuting space-time and
momentum.\medskip\\ Similarly, from formulae (\ref{r310}) and
(\ref{r318}) we obtain the commutation relations for the linear
basis of TD-{\it twisted phase space} $\Pi_\kappa(\beta) =
{\cal{X}}_\kappa(\beta)>\!\!\!\triangleleft\,
{\cal{P}}_\kappa$\begin{equation}
\begin{array}{lll} [ x_0 , x_i ] \ = & -\frac{i}{\kappa c} x_i\, ,
\qquad\qquad & [ x_i , x_j ] \ = \ 0\, ,\medskip\cr [ x_i , p_0 ]
\ = &  0\, , & [ x_0 , p_i ] \ = \  \frac{i}{\kappa c} p_i +
\beta_{j i} x_j\, ,\medskip\cr [ x_0 , p_0 ] \ = & - i\, , & [
p_\mu , p_\nu ] \ = \ 0\, ,\bigskip\cr [ x_i , p_j ] \ = &
i\,\delta_{i j}  \, .&\end{array}\label{r323}\end{equation}We see
that commutators are given by linear combinations of $p_i\,, x_i$
therefore one can regard these formulae as defining phase space
(not an algebra) in the classical sense. Also in this case we can
consider the limit $\kappa\to\infty, \beta=const.$ (see
(\ref{r311})) of phase space $\Pi_\kappa(\beta) \to
\Pi_\infty(\beta)$ given by the non-vanishing
commutators\begin{equation}[ x_0 , p_0 ] \ = \ -i\, ,\quad [ x_i ,
p_j ] \ = \ i\,\delta_{i j}\, ,\quad [ x_0 , p_i ] \ = \
i\,\beta_{j i} x_j\, ,\label{r324}\end{equation}with commuting
space-time and momentum.\medskip\\It turns out that both phase
spaces $\Pi_\infty(\alpha)$ and $\Pi_\infty(\beta)$ can be
realized by the standard position and momentum operators
$\hat{x}_\mu ,\, \hat{p}_\nu$ satisfying the Heisenberg
commutation relations $[ \hat{x}_\mu , \hat{p}_\nu ] = -i\,g_{\mu
\nu}$\begin{equation} x_i \ = \ \hat{x}_i e^{\alpha \hat{x}_0}\,
,\quad x_0 \ = \ \hat{x}_0\, ,\quad p_\mu \ = \ \hat{p}_\mu\,
,\qquad \textrm{for} \qquad \Pi_\infty(\alpha)\, ,
\label{r325}\end{equation}and assuming $\beta_{i j} =
\beta\,\delta_{i j}$\begin{equation} x_\mu \ = \ \hat{x}_\mu\,
,\quad p_0 \ = \ \hat{p}_0\, ,\quad p_i \ = \ \hat{p}_i - \beta
\hat{p}_0\, \hat{x}_i\, ,\qquad \textrm{for} \quad
\Pi_\infty(\beta)\, .\label{r326}\end{equation}We notice that both
algebras ${\cal{X}}_\kappa$ and ${\cal{P}}_\kappa$ possess the
Hopf structure therefore one can also consider a left action of
space-time algebra on the momentum algebra $ \triangleright :
{\cal{P}}_\kappa\otimes {\cal{X}}_\kappa\rightarrow
{\cal{X}}_\kappa$ formally a changing the position and momentum
generators $ x\leftrightarrow p$. It corresponds in quantum
mechanical language to the exchange of momentum for positions
representation. In this case one can define a phase space as the
cross product algebra ${\cal{P}}_\kappa
>\!\!\!\triangleleft \,{\cal{X}}_\kappa$ (see \cite{LN1}) with slightly
different cross commutation relations. However, we do not consider
twisting of this kind of phase space.

\section{Duality for twisted D=2 space-time}
Considering the formulae (\ref{r27c}) we notice that different
choices of coproducts $\Delta(x)$ or $\Delta(p)$ provide changes
in duality relations $< p^m\,q^n , x >$ and $< p , x^k\,y^l >$,
respectively. Therefore, we can expect the modified duality
relations $< p_i^m\,p_0^n , x_j >$ between four-momentum $p_\mu$
in the bicrossproduct basis (\ref{r24}) and SD-twisted or
TD-twisted space-time given by relations (\ref{r38}) or
(\ref{r310}). We find these twisted duality relations in the case
of two dimensional (D=2) space-time applying tensor methods to
twisted coproduct. This simplification is convenient because of
tedious calculations, however, one can immediately generalize the
obtained results to the four-dimensional space-time.
\subsection{SD-twisted space-time}
In the case of the two dimensional SD-twisted space-time relations
(\ref{r38}) take the following form (we assume $c=1$ in order to
simplify the notation)\\-- {\it space-time}
\begin{eqnarray}
[ x_0 , x ] &=& < f(p_0) , x_0 > \,x\, ,\label{401}\\ \Delta (x_0)
&=& x_0\otimes 1 + 1\otimes x_0\, ,\qquad \Delta(x) \ = \ x\otimes
e^{\alpha x_0} + e^{\alpha x_0}\otimes x\,
,\label{402}\end{eqnarray}-- {\it momentum space}
\begin{eqnarray}[ p_0 , p ] &=& 0\, ,\label{403}\\
 \Delta(p_0) &=& p_0\otimes 1 + 1\otimes p_0\,
,\qquad \Delta(p) \ = \ p\otimes 1 + f(p_0)\otimes p\,
,\label{404}\end{eqnarray} where we denote $f(p_0) =
\exp\left(-{p_0/\kappa}\right)$,\\-- {\it duality relations} are
given by (\ref{r27a})-(\ref{r27c}) for two dimensional case ($\mu
, \nu = 0, 1$).\bigskip\\One can also describe these duality
relations in terms of a nonlinear transformed basis of
${\cal{X}}_\kappa$ (\ref{r3d7})
\begin{equation} s(x_0) \ = \ x_0\, ,\qquad s(x_1) = s(x) = x
e^{\alpha x_0}\, ,\label{405}\end{equation}and using a trivial
coproduct $\Delta_0(x_\mu)$ (\ref{r210b})
\begin{eqnarray} &<p_\mu , s(x_\nu) > = i g_{\mu \nu}\, ,
& g_{\mu\nu} = (1, -1)\, ,\nonumber\\ & < p q , s > = < p\otimes q
, \Delta_0(s) >\, , & < p , s s' > = < \Delta(p) , s\otimes s' >\,
,\label{406}\end{eqnarray}for $s, s'\in {\cal{X}}_\kappa,\, p,
q\in {\cal{P}}_\kappa$.

\noindent Taking into account the coproduct relations (\ref{402})
and (\ref{404}) we can immediately generalize the relations
(\ref{r27b}) and obtain
\begin{equation}
< p^m_\mu , 1 > \  = \ < 1 , x^m_\mu > = \delta_{m 0}\, ,\qquad m
= 0 , 1, 2 ,\dots\, .\label{408}\end{equation}In order to find
other duality relations we apply the useful relation which
expresses the coassociativity of coproduct (see
\cite{M})\begin{eqnarray}< p_\mu^m , x_\nu^k > &=& <
\Delta^{(k-1)}(p_\mu^m) , x_\nu^{\otimes k} > \ = \ <
\left(\Delta^{(k-1)} (p_\mu)\right)^m , x_\nu^{\otimes k} > \ =
\nonumber\medskip\\ &=& < p_\mu^{\otimes m} , \Delta^{(m-1)}
(x_\nu^k) > \ = \ < p_\mu^{\otimes m} , \left(\Delta^{(m-1)}
(x_\nu)\right)^k
>\, ,\label{409}\end{eqnarray}where\begin{equation}\Delta^{(n)} \ = \
(I^{\otimes (n-1)}\otimes \Delta ) \Delta^{(n-1)} \ = \
(\underbrace{1\otimes\cdots\otimes1}_{(n-1)-times}\otimes \Delta)
\Delta^{(n-1)}\,
,\label{410}\end{equation}\begin{equation}x_\nu^{\otimes k} \ = \
\underbrace{x_\nu\otimes x_\nu\otimes\cdots \otimes x_\nu}_{k -
times}\, ,\qquad p_\mu^{\otimes m} \ = \ \underbrace{p_\mu\otimes
p_\mu\otimes \cdots \otimes p_\mu}_{m - times}\,
.\label{411}\end{equation}In particular\begin{eqnarray}
\Delta^{(m-1)} (x) &=& \sum_{i=1}^{m} x_i^{(m-1)}\, , \qquad\quad
x_i^{(m-1)}\ =\ (e^{\alpha x_0})^{\otimes (i-1)}\otimes x\otimes
(e^{\alpha x_0})^{\otimes (m-i)}\, ,\label{412}\\
\Delta^{(m-1)}(x_0)& =& \sum_{i=1}^{m} (x_0)_i^{(m-1)}\, ,\qquad
(x_0)_i^{(m-1)}\ =\ I^{\otimes (i-1)}\otimes x_0\otimes I^{\otimes
(m-i)}\, ,\label{413}\\ \Delta^{(k-1)} (p)& =& \sum_{i=1}^{k}
p_i^{(k-1)}\, ,\qquad\quad\; p_i^{(k-1)}\ =\ f^{\otimes
(i-1)}\otimes p\otimes I^{\otimes (k-i)}\, ,\label{414}\\
\Delta^{(k-1)} (p_0)& =& \sum_{i=1}^{k} (p_0)_i^{(k-1)}\,
,\qquad\;\, (p_0)_i^{(k-1)}\ =\ I^{\otimes (i-1)}\otimes
p_0\otimes I^{\otimes (k-i)}\, . \label{415}\end{eqnarray}Using
the coproduct formulae (\ref{402}) for space-time variables,
duality relations and relations (\ref{412}-\ref{413}) we
obtain\begin{eqnarray} < p_0^n , x_0 > =& \delta_{n 1} <p_0 , x_0
>\, ,\quad & < p^m , x_0 > \ = \ 0\, ,\label{416}\\ <p_0^n , x > =&
0\, ,\qquad\qquad\qquad & < p^m , x > \ = \ \delta_{m 1} < p , x
>\, .\label{417}\end{eqnarray}For instance\begin{eqnarray} < p^m ,
x_0 > &=& < p^{\otimes m} , \Delta^{(m-1)} ( x_0) > \ = \cr &=&
\sum_{i=1}^{m}< p\otimes \cdots \otimes p , I^{\otimes
(i-1)}\otimes x_0 \otimes I^{\otimes (m-i)} > \ =\cr &=& m < p , 1
>^{m-1} < p , x_0> \ = \ 0\, .\nonumber\end{eqnarray}
Analogously, using the momentum coproduct (\ref{404}) and formulae
(\ref{414})-(\ref{415}) we derive
\begin{eqnarray} < p_0 , x_0^l > = & \delta_{l 1} <p_0 , x_0
>\, ,\quad & < p , x_0^n > \ = \ 0\, ,\label{418}\\ <p_0 , x^k > =
& 0\, ,\qquad\qquad\qquad  & < p , x^k > \ = \ \delta_{k 1} < p ,
x >\, .\label{419}\end{eqnarray} The relations
(\ref{416})-(\ref{417}) or (\ref{418})-(\ref{419}) can be
generalized to the following form\begin{eqnarray} < p_0^n , x_0^l
> &=& n!\,<p_0 , x_0>^n \delta_{n l}\, ,\quad < p^m , x_0^l > \ =
\ \delta_{m 0} \delta_{l 0}\, ,\label{420}\\ < p^m , x^k > &=&
m!\,<p , x >^m \delta_{m k}\, ,\quad < p_0^n , x^k > \ = \
\delta_{n 0}\delta_{k 0}\, .\label{421}\end{eqnarray}From
(\ref{415}) and the trivial form of coproduct (\ref{404}) we
compute for instance\begin{eqnarray}< p_0^n , x_0^l > &=& <
\Delta^{(l-1)}(p_0^n) , x_0^{\otimes l} > \ = \cr &=& <
\left(\sum_{i=1}^{l}(p_0)_i^{(l-1)}\right)^n , x_0\otimes \cdots
\otimes x_0 > \ = \cr &=&  n! < p_0^{\otimes n} , x_0^{\otimes l}
> \ = \ n! < p_0 , x_0 >^l \delta_{n l}\, .\nonumber\end{eqnarray}
We would like to stress that the duality relations
(\ref{420})-(\ref{421}) do not depend on SD-twisting
transformation and they have the same form in both,
$\kappa$-deformed and non-deformed cases.\medskip\\Let us derive
the relation which depends on the twisting parameter. Using
(\ref{420})-(\ref{421}) and coproduct formula (\ref{402}) for
$\Delta(x)$ we find

\begin{eqnarray} < p_0^n p^m , x > &=& < p_0^n\otimes p^m ,
\Delta(x)> \ = \ < p_0^n\otimes p^m , x\otimes e^{\alpha x_0} +
e^{\alpha x_0}\otimes x >\ =\cr &=& \delta_{m 1} < p , x > < p_0^n
, e^{\alpha x_0} >\ =\ \delta_{m
1}\,\sum_{k=0}^{\infty}\frac{\alpha^k}{k!} < p , x > < p_0^n ,
x_0^k >\ =\cr &=& \alpha^n \delta_{m 1} < p , x
> < p_0 , x_0 >^n\, ,\label{422}\end{eqnarray}
\noindent It is convenient to use instead the power function
$p_0^n\,$, its generating function $g(p_0)=e^{\xi p_0}$. Then we
obtain
\begin{equation} < p^m\,e^{\xi p_0} , x > \ = \ \delta_{m 1}\,< p
, x > e^{\alpha \xi <p_0,x_0>}\, ,\label{424}\end{equation}or
using (\ref{414}) we find a more general formula\begin{equation} <
p^m\,e^{\xi p_0} , x^k > \ = \ k! (< p , x >)^k \delta_{m k
}\,e^{k \alpha \xi <p_0 , x_0>}\, f^{\frac{1}{2} k(k-1)}(\alpha
<p_0 , x_0>)\, .\label{425}\end{equation}From (\ref{420}) we can
easy calculate the following duality relation
\begin{eqnarray}< p^m\,e^{\xi
p_0} , x_0^l > &=&
\sum_{s=0}^{l}\left(\begin{array}{c}l\\s\end{array}\right) <
p^m\otimes e^{\xi p_0} , x_0^{l-s}\otimes x_0^s >\ = \cr &=&
\sum_{s=0}^{l}\left(\begin{array}{c}l\\s\end{array}\right) < p^m ,
x_0^{l-s}> < e^{\xi p_0} , x_0^s >\ =\cr &=& \delta_{m 0} < e^{\xi
p_0} , x_0^l >  \ = \ (\xi <p_0 , x_0
>)^l \delta_{m 0}\, ,\label{426}\end{eqnarray}and finally we
find\begin{eqnarray}&&< p^m\,e^{\xi p_0} , x^k x_0^l > \ = \ <
\Delta(p^m\,e^{\xi p_0}) , x^k\otimes x_0^l >\ = \cr &&= \
\sum_{s=0}^{m} \left(\begin{array}{c}m\\s\end{array}\right)<
p^{m-s} f^s(p_0)\,e^{\xi p_0}\otimes p^s e^{\xi p_0} , x^k\otimes
x_0^l
>\ =\cr &&= \ \sum_{s=0}^{m}\left(\begin{array}{c}m\\s\end{array}\right)
< p^{m-s} f^s(p_0) e^{\xi p_0} , x^k > < p^s e^{\xi p_0} , x_0^l
>\ = \cr &&=  \ (\xi<p_0 , x_0>)^l < p^m\,e^{\xi p_0} , x^k
>\ =\cr && = k! (<p , x>)^k \delta_{m
k} (\xi<p_0 , x_0>)^l e^{k \alpha \xi <p_0,x_0>}
f^{\frac{1}{2}k(k-1)}(\alpha<p_0,x_0>)\,
,\label{427}\end{eqnarray}or equivalently (expanding the
generating function in powers of $p_0^n$)\begin{equation} < p^m
p_0^n , x^k x_0^l > \ = \ k! n! (<p , x>)^k (<p_0 , x_0>)^l
\delta_{m k} \frac{(i k\alpha)^{n-l}}{(n-l)!}\,
f^{\frac{1}{2}k(k-1)}(\alpha<p_0,x_0>)\,
.\label{428}\end{equation}In the limit $\alpha \to 0$ of the
twisting parameter we obtain the duality relations (\ref{r29}).
Therefore, we see that the twisting map destroys some orthogonal
duality relations i.e. roughly speaking twisting changes the
orthogonal basis of dual momentum and space-time algebras to a
non-orthogonal one.

\noindent Because twisted space-time is a non-commuting algebra,
one can also consider a polynomial basis with opposite ordering of
space and time variables i.e. left-time ordered polynomials. In
order to find the duality relations for this case, we would like
to stress that space-time commutation relations (\ref{r213}) are
related to the momentum coproduct in the bicrossproduct basis and
do not depend on the twisting map, therefore we can use these
relations to derive duality relations for the opposite ordering.
Taking into account (\ref{r27a}) and the explicit form of the
function $f(p_0)$ we obtain the SD-{\it twisted duality
relations}\begin{eqnarray}< p^m\,e^{\xi p_0} , x^k x_0^l > & = &
k! (-i)^k (i \xi)^l \delta_{m k} e^{i k \alpha \left(\xi +
\frac{1-k}{2\kappa}\right)}\, ,\label{431}\\< p^m e^{\xi p_0} ,
x_0^l x^k
> &=&  k!\,(-i)^k \delta_{m k}  e^{i k \alpha \left(\xi +
\frac{1-k}{2\kappa}\right)} < e^{-\frac{k p_0}{\kappa}} , (x_0 +
i\xi)^l >\, ,\label{432}\end{eqnarray}or equivalently
\begin{eqnarray}< p^m p_0^n , x^k x_0^l >  &=&  k!\,\delta_{k m} (-i)^k
e^{-\frac{i\alpha}{2\kappa} k(k-1)} <(p_0 + ik\alpha)^n , x_0^l
>\, ,\label{433}\end{eqnarray} and a non-vanishing duality relation only for $n\leq l$
\begin{eqnarray}< p^m p_0^n , x_0^l x^k >  &=& k!\,\delta_{k m}
(-i)^k e^{-\frac{i\alpha}{2\kappa} k(k-1)} < e^{-\frac{m
p_0}{\kappa}} (p_0 + im\alpha)^n , x_0^l >\,
.\label{434}\end{eqnarray} The duality relations
(\ref{433})-(\ref{434}) one can also describe in terms of a
transformed basis $s(x_\mu)$ (\ref{405}) as follows
\begin{eqnarray}< p^m p_0^n , x^k x_0^l >  &=& < p^m\otimes p_0^n
, \Delta_0\left(s^k(x) s^l(x_0)\right) >\, ,\label{435a}\\< p^m
p_0^n , x_0^l x^k > &=& < p^m\otimes p_0^n ,
\Delta_0\left(s^l(x_0) s^k(x)\right)
 >\, .\label{435b}\end{eqnarray}

\subsection{TD-twisted space-time}
Similarly, in the two-dimensional case the TD-twisted space-time
(\ref{r310}) is given by the relations\begin{eqnarray}[ x_0 , x ]
&=& < f(p_0) , x_0
> x\, ,\cr \Delta( x_0 ) &=& x_0\otimes 1 + 1\otimes x_0 + \beta\,x\otimes x\,
,\qquad \Delta( x ) \ = \ x\otimes 1 + 1\otimes x\,
,\label{436}\end{eqnarray}and additional formulae
(\ref{403})-(\ref{404}) describing the momentum space and duality
relations. Also in this case the relations (see
(\ref{420})-(\ref{421}))
\begin{equation} < p_0^n , x_0^l
> = n!\,<p_0 , x_0>^n \delta_{n l}\, ,\quad <
p^m , x^k > = m!\,<p , x >^m \delta_{m k}\, ,
\label{437}\end{equation} are valid and the remaining ones are
changed. It is easy to observe that the form of coproduct
$\Delta(x_0)$ (\ref{436}) implies a vanishing duality relation for
any odd power of the momentum
\begin{eqnarray}<p^m , x_0 > &=& < p^2 ,
x_0 > \delta_{m 2} \ = \ \beta <p , x>^2 \delta_{m 2}\, ,\cr <
p^{2k+1} , x_0^l > &=& 0\quad \Leftrightarrow\quad < \sinh(\xi p)
, x_0^l > \ = \ 0\, .\label{438}\end{eqnarray}Let us derive the
duality relations for even power of the momentum. Using the
generating function we obtain
\begin{eqnarray} \qquad< e^{\xi p} , x_0^l > &=& < \cosh(\xi p) , x_0^l >
\ = \  < \Delta^{(l-1)} \exp(\xi p) , x_0^{\otimes l} >\ = \cr &=&
< \exp\left( \xi \Delta^{(l-1)}(p)\right) , x_0^{\otimes l}
>=< \exp\left( \xi \sum_{i=1}^{l} p_i^{(l-1)}\right) ,
x_0^{\otimes l} >\ =\cr &=& \prod_{i=1}^{l} < \cosh\left( \xi
p_i^{(l-1)}\right) , x_0^{\otimes l} >\,
.\label{439}\end{eqnarray}The non-vanishing duality relations are
implied by the square power of the momentum variable (\ref{438}),
therefore we can use the expansion\begin{equation}\cosh\left( \xi
p_i^{(l-1)}\right) \ \sim \ I^{\otimes l} + \frac{1}{2} \xi^2
\left( p_i^{(l-1)}\right)^2\, ,\label{440}\end{equation}and we
get\begin{eqnarray}&&<\cosh(\xi p) , x_0^l > \ = \ \prod_{i=1}^{l}
< \left(I^{\otimes l} + \frac{1}{2} \xi^2 \left(
p_i^{(l-1)}\right)^2\right) , x_0^{\otimes l} >\ =\cr&& = \
\delta_{l\,0} \ + \ \sum_{k=1}^{l} \frac{1}{2^k} \xi^{2k}
D_{k}^{l}\, ,\label{441}\end{eqnarray} where\begin{equation}
D_k^{l}\ \equiv\ D_k^l(\kappa , \beta) \ = \ \sum_{i_1\neq
i_2\neq\cdots i_{k-1}}^{l-1} < (p_{i_1}^{(l-1)})^2
(p_{i_2}^{(l-1)})^2\cdots (p_{l}^{(l-1)})^2 , x_0^{\otimes l} >\,
,\label{442}\end{equation}satisfying\begin{equation} D_k^l(0 , 0)
\ = \ D_k^l( \kappa , 0 ) \ = \ 0\, ,\label{442a}\end{equation}and
comparing the appropriate left and right terms in (\ref{441}) we
find nonvanishing relations for $m\leq l$\begin{equation}< p^{2m}
, x_0^l
> \ = \ \frac{(2m)!}{2^m} D_m^{l}(\kappa , \beta)\, .
\label{443}\end{equation} Now, we can derive a more general
formula\begin{eqnarray}< e^{\xi p_0}\,p^m , x_0^l > &=& < e^{\xi
p_0}\otimes p^m , \Delta^l(x_0) > \ = \ < e^{\xi p_0}\otimes p^m ,
\Delta_0^l(x_0) >\ =\cr &=& < e^{\xi p_0}\otimes p^m , (x_0\otimes
1 + 1\otimes x_0)^l >\ =\cr &=& \sum_{s=0}^{l}
\left(\begin{array}{c}l\\s\end{array}\right)< e^{\xi p_0}\otimes
p^m , x_0^{l-s}\otimes x_0^s >\ =\cr &=& \sum_{s=0}^{l}
\left(\begin{array}{c}l\\s\end{array}\right) < e^{\xi p_0},
x_0^{l-s} > < p^m , x_0^s
>\, ,\label{444}\end{eqnarray}or expanding we get non-vanishing duality relations
for $n \leq l$\begin{equation}< p_0^n p^m , x_0^l> \ = \
\frac{l!}{(l-n)!} <p_0 , x_0 >^n < p^m , x_0^{l-n} >\,
.\label{445}\end{equation} Finally, using this relation and
(\ref{437}) and coproduct formula (\ref{404}) we derive the
general duality relations for the TD-{\it twisted space-time}
\begin{eqnarray}< p^m p_0^n , x^k x_0^l
> &=& < \Delta^m(p) \Delta^n(p_0) , x^k\otimes x_0^l > \cr &=&
\frac{m!\,n! <p , x>^k <p_0 , x_0>^n}{(m - k)! (l - n)!} < p^{m -
k} , x_0^{l - n} >\, ,\label{446}\end{eqnarray}or equivalently we
get a nonvanishing duality relations for $m-k=2s\,, s\leq l-n\,,
(s=0,1,2,\dots)$
\begin{eqnarray}< p^m p_0^n , x^k x_0^l > &=& <
\Delta^m(p) \Delta^n(p_0) , x^k\otimes x_0^l > \cr
 &=& \ \frac{m!\,n! <p , x>^k <p_0 , x_0>^n}{(m
- k)! (l - n)!}\, \frac{(2s)!}{2^s} D_s^{l-n}(\kappa , \beta)\,
.\label{447}\end{eqnarray}In the limit $\beta\to 0$ using
(\ref{442a}) we obtain the duality relation (\ref{r29}). Therefore
we see that TD-twisting appears as an additional term to the
conventional duality relations (\ref{r29}). One can also easily
find the duality relations for the opposite space-time ordering
using formula (\ref{r213}) but they have rather complicated
form.\medskip\\Also in this case one can describe the duality
relations (\ref{446}) in terms of nonlinearly transformed basis
(\ref{r3d8})
\begin{equation}< p^m p_0^n , x^k x_0^l > \ = \ < p^m\otimes p_0^n ,
\Delta_0\left(t^k(x) t^l (x_0)\right) >\,
.\label{448}\end{equation}

\subsection{Space-time and phase-space; beyond $\kappa$-deformed
framework} One can give up the assumption of a commutative dual
momentum space (required by $\kappa$-deformed symmetry) and
consider a noncommuting fourmomentum algebra dual to the
space-time algebra ${\cal{X}}_\kappa$ defined by
(\ref{r210a})-(\ref{r210b}). In such a case we deal with
nonsymmetric space-time coproduct. As an example we shall discuss
space-time algebra obtained by the Jordanian twist \cite{Ogiev}
(see also \cite{Jord}) of two dimensional space-time
${\cal{X}}_\kappa$. First we notice that the relations
(\ref{r210a})-(\ref{r210b}) for the case D=2 describe a Lie
algebra $B(2)$ isomorphic to a Borel subalgebra of $sl(2)$
therefore, we can apply the Jordanian twisting function (with
$\xi$ as a twisting parameter) in the form \cite{Jord}
\begin{equation} F_J(\kappa , \xi) \ = \ e^{i \kappa c x_0\otimes \sigma_\xi(x)}
\ = \  e^{i \kappa c x_0\otimes \ln (1+ \xi x)}\,
,\label{449}\end{equation}to the trivial coproduct
$\Delta_0(x_\mu)$, ($\mu=0,1,\, x_1=x$)
\begin{equation}\Delta_J(x_\mu) \ = \ F_J(\kappa , \xi)
\Delta_0(x_\mu) F^{-1}_J(\kappa , \xi)\,
.\label{450}\end{equation}This coproduct defines the following two
dimensional Jordanian twisted space-time ${\cal{X}}^J_\kappa(\xi)$
\begin{eqnarray}&& [ x_0 , x ] \ = \ -\frac{i}{\kappa c}\,x\,
,\qquad\qquad \epsilon(x_\mu) \ = \ 0\,
,\label{451}\\&&\Delta_J(x_0) \ = \ x_0\otimes e^{-\sigma_\xi(x)}
+ 1\otimes x_0 \ = \ \Delta_0(x_0) - \xi x_0\otimes x (1+\xi
x)^{-1}\, ,\label{452}\\&&\Delta_J(x) \ = \ x\otimes 1 +
e^{\sigma_\xi(x)}\otimes x \ = \ \Delta_0(x) + \xi x\otimes x\,
,\label{453}\\&&S_J(x_0) \ = \ - x_0 e^{\sigma_\xi(x)} \ = \ -x_0
(1 + \xi x)\, ,\label{454}\\&&S_J(x) \ = \ - x e^{-\sigma_\xi(x)}
\ = \ -x (1 + \xi x)^{-1}\, .\label{455}\end{eqnarray}From the two
dimensional version of duality relations (\ref{r27a})-(\ref{r29})
one can find a dual momentum algebra ${\cal{P}}^J_\kappa(\xi)$.
Because there is an isomorphism between the Hopf algebras
${\cal{X}}^J_\kappa(\xi)$ and its dual ${\cal{P}}^J_\kappa(\xi)$
\cite{Ogiev} given by $\sigma_\xi(x)\rightarrow p_0/\kappa c\, ,
\quad x_0\rightarrow p/\kappa c \xi$ therefore we can find the
defining relations ($\mu = 0,1,\, p_1=p$)
\begin{eqnarray}&&[ p_0 , p ] \ = \ i \kappa c \xi \left(1 -
e^{-\frac{p_0}{\kappa c}}\right)\, ,\qquad \epsilon(p_\mu) \ = \
0\, ,\label{456}\\&&\Delta(p_0) \ = \ p_0\otimes 1 + 1\otimes
p_0\, ,\qquad S(p_0) \ = \ - p_0\, ,\label{457}\\&&\Delta(p) \ = \
p\otimes 1 + e^{-\frac{p_0}{\kappa c}}\otimes p\, ,\qquad S(p) \ =
\ - e^{\frac{p_0}{\kappa c}}p\, .\label{458}\end{eqnarray} However
the form of the coproduct is the same as for the $\kappa$-deformed
momentum algebra (\ref{404}), the momentum algebra becomes
noncommuting.\medskip\\This pair of dual algebras we can extend to
a Jordanian phase space algebra $\Pi_\kappa^J(\xi)$ by the left
cross product construction $\Pi_\kappa^J(\xi) =
{\cal{X}}_\kappa^J(\xi)
>\!\!\!\!\triangleleft\,{\cal{P}}_\kappa^J(\xi)$ (\ref{r319}) and we
find the following commutation relations of the linear basis
\begin{eqnarray}&[ x_0 , x ] \ = \ -\frac{i}{\kappa c}\,x\,
,\qquad &[ p_0 , p ] \ = \ i \kappa c \xi \left(1 -
e^{-\frac{p_0}{\kappa c}}\right)\, ,\label{459}\\&[ x , p_0 ] \ =
\ 0\, ,&[ x_0 , p ] \ = \ i \left(\frac{p_0}{\kappa c} - \xi
x_0\right)\, ,\label{460}\\&[ x_0 , p_0 ] \ = \ -i\, , & [ x , p ]
\ = \ i (1 + \xi x)\, .\label{461}\end{eqnarray}In the limit
$\xi\to 0$ we obtain the standard $\kappa$-deformed phase space
\cite{LN1}.

\section{Final remarks}
It is worthwhile to notice, that one can describe the duality
relations between space-time and momentum algebras in terms of
linearly transformed momentum basis. This possibility is similar
to the description using the functions $s(x_\mu) , t(x_\mu)$. We
consider the case of SD-twisting in the two-dimensional case.
First we notice that the duality relations (\ref{433}) allow us to
define a linear (because of the bilinear form $< \ , \ >$)
transformation $\Phi_\alpha$ in the momentum algebra
${\cal{P}}_\kappa$ corresponding to the twist operation in the
space-time ${\cal{X}}_\kappa(\alpha)$ as follows\begin{equation}<
p^m p_0^n , x^k(\alpha) x_0^l(\alpha)> \ = \ < \Phi_\alpha(p^m
p_0^n) , x^k x_0^l > \ = \ < f_{m n}(\alpha) , x^k x_0^l >\,
,\label{51}\end{equation}where ($x(\alpha)\,, x_0(\alpha)$) are
space-time variables (\ref{401})-(\ref{402}) generating the
twist\-ed algebra ${\cal{X}}_\kappa(\alpha)$ and $x\,,x_0 \in
{\cal{X}}_\kappa$ (see (\ref{r210a})-(\ref{r210b})) or in explicit
form\begin{equation} \Phi_\alpha(p^m\,p_0^n) \ = \ f_{m\,
n}(\alpha) \ = \ ( p_0 + i m \alpha )^n p^m\,e^{-\frac{i\alpha}{2
\kappa} m(m-1)}\, .\label{52}\end{equation}In
particular\begin{eqnarray} \Phi_\alpha( p^m ) &=& f_{m\,0}(\alpha)
\ = \ p^m\,e^{-\frac{i\alpha}{2 \kappa} m(m-1)}\,
,\label{53}\medskip\\ \Phi_\alpha ( p_0^n ) &=& f_{0\,n}(\alpha) \
= \ p_0^n\, ,\label{54}\medskip\\ \Phi_\alpha( 1 ) &=& 1\,
,\label{55}\end{eqnarray}therefore, the action of $\Phi_\alpha$ on
the momentum space (generated linearly  by $p_0\,, p$) is trivial
because $p_0 =f_{0 1}(\alpha)\,, p=f_{1 0}(\alpha)$  and their
coproduct is given by (\ref{r26}). The action of $\Phi_\alpha$ on
the momentum algebra basis $p^m p_0^n$ allows us to extend this
transformation onto an arbitrary polynomial function of momentum
$\phi = \phi_{m\,n} p^m p_0^n$ in a natural way, by the
replacement $p^m p_0^n\rightarrow f_{m\,n}(\alpha)$. Thus, one can
consider the dual pair (${\cal{X}}_\kappa(\alpha),
{\cal{P}}_\kappa$) of the twisted space-time algebra and the
$\kappa$-deformed momentum algebra or equivalently the pair of a
algebras (${\cal{X}}_\kappa , {\cal{P}}_\kappa(\alpha)$) of
$\kappa$-deformed space-time and $\Phi_\alpha$-transformed
momentum algebra with the same duality relations (\ref{51}) in
both cases. The essential difference in both dual constructions
lies in their coalgebra sectors.\medskip\\ Our construction of the
SD-twisted phase space algebra $\Pi_\kappa(\alpha)$ (\ref{r321})
leans on the notion of the cross algebra where the multiplication
(\ref{r319}) depends on both coproducts (\ref{402}) and
(\ref{404}). Therefore, for the second pair of algebras
(${\cal{X}}_\kappa , {\cal{P}}_\kappa(\alpha)$) we obtain the
different phase space algebra although both pairs are equivalent
as far as the duality relations are concerned. In derivation of
phase space commutation relations we use only the first order
duality relations (for instance in the definition of the left
action (\ref{r317})) therefore, in the case of the pair
(${\cal{X}}_\kappa , {\cal{P}}_\kappa(\alpha)$) we obtain a
two-dimensional version of commutation relations (\ref{r321}) for
$\alpha=0$ i.e. the standard $\kappa$-deformed phase space. The
same conclusions one can obtain by considering TD-twisting
space-time.\medskip\\Therefore, one can find the linear
transformation of the momentum algebra which corresponds the twist
operation in the space-time algebra, this construction however
does not provide the twisted phase space.\medskip\\In Section 3 we
described two possible twistings (in space and time directions) of
the space-time algebra and derived corresponding phase
spaces.\medskip\\The duality relations obtained in Section 4 for
two-dimensional space-time and momentum algebras one can
immediately extend to the four-dimensional case. They describe
explicitly the effect of the twisting operation in the space-time
algebra.

\section*{Acknowledgements} The author wish to thank Anatol Nowicki
for inspiration, advice and support. I would like to thank the
referee for useful suggestions and comments.

\end{document}